# Security Issues in WBANs

**Muhammad Asam**[+]

Email: asim2k994@gmail.com
PhD Scholar

**Tauseef Jamal**[+]

Email: jamal@pieas.edu.pk
Associate Professor, PIEAS

**Aleena Ajaz**[+]

Email: aleena.ajazsh@gmail.com

**Zeeshan Haider**[+]

Email: luckier19@gmail.com
PhD Scholar

**Shariq Butt**[*]

Email: shariq2315@gmail.com
PhD Scholar

[+]*CoCoLabs PIEAS, Islamabad, Pakistan*
[*]*CoCoLab, University of Lahore, Lahore, Islamabad*

**Abstract----** **Wireless Body Area Network (WBAN) refers to short-range, wireless communications near or inside a human body. WBAN is emerging solution to cater the needs of local and remote health care related facility. Medical and non-medical applications have been revolutionarily under consideration for providing a healthy and gratify service to the humanity. Being very critical in communication from body it faces a lot of challenges which are to be tackled for the safety of life and benefit of the user.**

**There is variety of challenges faced by WBANs. WBAN is favorite playground for attackers due to its usability in various applications. This article provides systematic overview of main challenges in WBANs in security perspectives.**

Keywords: Wireless Body Area Networks; Denial of Service Attacks; Resource Management;

## 1. Introduction

Wireless communication brought numerous benefits to our society. Technology up gradation has made this communication possible by the help of 4G, LTE-A, 5G and so on. Recently, Machine to Machine (M2M) communication has been a favorite area of research in past few decades. Communication between machines and the human was next destination. Low power, lightweight and miniature physiological sensors has made it possible to connect them to form a Body Area Network (BAN). This connection is supplemented by the wireless technology and the WBAN is formed. WBAN comprises multiple sensors to sample, process and communicate vital sign like heart beat rate, vascular blood pressure and/or blood oxygen saturation. Same can be done by the sensors for environmental parameters like location, temperature, humidity and light.

Communication network in WBAN can be divided into two major tiers, one is the communication between the sensors and the second is the distribution network [1]. Three-tier architecture is mostly agreed upon by inserting another layer of communication between WBAN coordinator and WBAN gateway or sink node.

First tier of WBAN architecture is realized by body sensor units, which are placed outside or inside of human body. These sensors are responsible for detecting the physiological data signals, converting the signals to digital form and then transmitting through wireless media. This communication is also referred as intra-BAN communication [1]. Second tier comprised of personal server units. These units get data from sensors and process it. This tier formats the processed results to convey to the upper, third tier if necessary. Communication with both the first and third tier is done wirelessly. This communication is also known as inter-BAN communication. Third tier comprises of user machines, where end users are data experts who can take some decision, or can conclude some results from this data. This inference may be about someone's health in hospital or at home. It may be sending some caretaker or ambulance to the patient [2].

IEEE 802.15.6 standard specifies the wireless communications near body or inside the body. This standard considers effects on portable antennas due to the presence of a person (varying with male, female, skinny, heavy, etc.), radiation pattern shaping to minimize the Specific Absorption Rate (SAR) into the body, and changes in characteristics as a result of the user motions. In the following, we described some of its applications.

As the WBAN is closely attached to acquire the body parameters so its most favorite applications are in medical field. ECG, pulse oximeter and heart beat sensors can be attached the patient's body to monitor irregularities in heart. Cancer can be detected by using nitric oxide emission sensors. Similarly, allergic sensors can be used to detect the allergic agents in the air to report to the patient or physician. Bio-sensors are implanted to monitor Glucose level and inject insulin automatically when the glucose level is at a certain threshold.

In order to measure the physiological data of the patient ambient sensors can be placed at home. This data is stored or transmitted to a control unit/healthcare center at periodic intervals. This helps the patients to stay at home and get continuous healthcare support without visiting the hospital. Moreover, these sensors, placed on the patient's body, will raise an alarm or urgent notification to the nearby healthcare center in case of any emergency.

The remainder of this paper is organized as follows: next section outlines the main challenges in WBANs. Section 3 discusses security issues, while Section 4 presents the summary and discussion. Finally Section 5 provides the conclusion and future work.

**2. Main Challenges in WBANs**

Effectiveness of the WBAN is important from both patients and healthcare perspective. As the time passes, challenges to the emerging technologies increases along with the advancements. There is variety of challenges faced by WBAN. We have classified these challenges in six major classes such as energy, mobility, security and communications (i.e., networking, QoS and cooperation), as shown in taxonomy provided in Fig. 1. Security is the major issue need to be tackle in parallel with any of other issues.

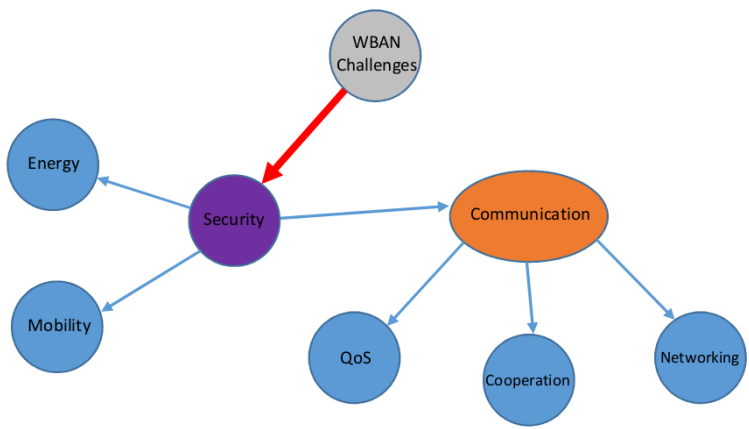

Fig. 1 WBAN main Challenges

As a rough estimate, weight of the battery is directly proportional to its power. So we may not increase the weight to increase the power as it is to be carried out by the human body and the case is more severe if it is to be implanted inside.

WBAN provides two major advantages, i.e., portable monitoring and location independence. Regardless of the application, these are the key factors due to which WBAN is potential candidate in many venues. However, these two advantages put specific limitation i.e., mobility. Mobility can pose serious problem in some application, e.g., in E-Health care even posture do effect the communication.

Quality of Service (QoS) is the requirements fulfilled by system as requested by the users. For more life critical system, timeliness may be the parameter for the quality. System, that cannot fulfill the said requirement, falls short of providing the QoS. Same is true for other factors like bandwidth, latency, jitter, robustness, trustworthiness, adaptability [3].

The major networking limitation come from limited resource devices and shared medium [4]. MAC layer is very important since it communicate with next hop and access the medium. Therefore, collisions, contention and resource blockage could be handled in case of efficient MAC layer schemes. Hence, backoff algorithms, Carrier Sence Multiple Access (CSMA), contention windows and handshaking need attention while devising MAC protocol for WBANs specially when there is ad-hoc connectivity.

As the size of network grows, it mainly affects the routing protocols performance and throughput of the network. Bandwidth utilization also suffers from links sharing in each connected node; this is one of major cause of slow routing in homogenous channelization.

**3. Security in WBAN**

In any network, communication data is of worth importance. In case of WBAN, it becomes more critical as it has been connected to the physical system. These communication channels are very much visible to the attacker and if not securely implemented, the attacks like eavesdroping on traffic, message injection, message replay or spoofing can occur. Upon successful attack, such actions not only invade privacy but may lead to catastrophic situation [5]. As reported in Healthcare IT news in February, 2014, hackers accessed a server from a Texas healthcare system, compromising the protected health information of some 405,000 individuals, which was one of the biggest HIPAA security breaches. Even worse, it was demonstrated that implantable cardiac devices can be wirelessly compromised [6].

Security measures are necessary to protect the users from potential risks. Security architecture for WBAN is more challenging than other networks. Efficiency, scalability and usability are performance requirements for the security architecture in WBAN. Regardless of the architecture of WBAN, we can coarsely divide the communication into two parts, internal communication between WBAN and external communication between WBAN and external users.

Security requirement in internal communication includes authenticity, confidentiality, integrity and availability of data and resources. Data authenticity means that data is coming from the claimed source. An attacker may inject bogus data into the WBAN. Public key cryptography schemes are used for data authenticity. Data confidentiality leads to information disclosure to unauthorized entities. Encryption is also done to achieve this. Data integrity is achieved through Message Authentication Code (MAC) or by the help of hashed MAC. Integrity is made sure by doing the reverse process of generating the authentication codes. Data and resource availability is the most pervasive security requirement for WBAN.

Due to its criticality of the physical system in WBAN, availability of data and resources is necessary. Denial of Service (DoS) attack is the favorite place for the attackers over here.

Along with completing the security requirement, WBAN protocols must be efficient enough to fulfil its desired mandate. In [7], the authors suggested secure and reliable routing framework for WBAN. They demonstrated that it can significantly counter the data injection attacks.

Utility of WBAN in healthcare system may include the self-monitoring patients, network service provider for data transmission, application support and local/remote personnel who offer medical services. Considering the privacy and significance of patient-related data and medical messages, WBAN may suffer threats such as message modification and unauthorized access. It is desirable that proper security mechanism should be considered for securing the communication between WBAN and external users, where each user must prove their authenticity and then access the data according to their privileges.

WBAN is favorite playground for attackers due to its usability in various applications. One of the classifications of attack is four part communication implementation stack namely PHY layer, MAC layer, network layer and transport layer attacks.

Being the radio frequency based, PHY is more prone to attacks like tempering and jamming. In jamming attack, attacker transmits radio signal of random frequency. This signal interferes with the other sensor signals. Eventually, node in the range of the attacker cannot communicate message and become isolated. In tampering attack, the cryptographic keys and even program code can be tampered.

MAC is dealing with the frame detection, multiplexing and channel accesses. Collision attack at this layer may cause in exponential rise in back-off packet in certain protocols. MAC schemes can be interrupted at this layer to cause unfairness attack. Continuous transmission of corrupted packets may result in DoS.

In WBAN, routing is carried through nodes coordination. A compromised node in a network can spoof, alter or replay the routing facts for the network. Sometime, the attacker may selectively route the packets in the network causing selective forwarding attack. A malicious node may attract all the traffic in the network to itself by claiming it to be the best coordinator in the network. It can do alteration with the data received once it is recognized as best data exchange. A single node may pose to have multiple network identities leading to Sybil attack.

An attacker may send a hello message powerful enough to be selected to route messages, leading to hello flood attack. An attacker may send many requests to establish the connection to use its all resources, causing flooding attack. This results in restricting making legitimate connections of the nodes. An attacker in de-synchronization attack sends fake control flags or sequence number to both nodes in an active connection.

### 3.1. Cooperation between Nodes

When the intermediate nodes help source destination pair in communication, the cooperation occurs. The intermediate nodes may refer as helper or relay as shown in Fig. 2.a [7].

Cooperation offers a good solution for many of the limitations in WBAN such as distance, mobility, coverage and mitigating attacks. Fig. 2.b shows that with cooperation we can reconstruct the path in case of an attack. However, introduction to cooperation may leads to additional threats, blockage and overhead, which limit its benefits and affects the resource management [8]. Therefore, communication is not reliable when cooperation is used.

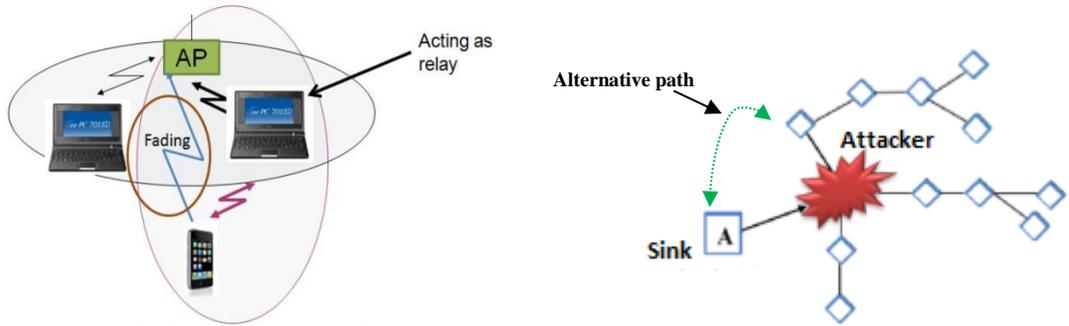

Fig. 2   (a) Cooperative Networking System          (b) Attack Mitigation via Cooperation

Communication via relay can add additional interference as well. Therefore, cooperation communication protocols must be devised in way to incur less overhead and interference. Another issue with cooperation is relay selection. Normally relays are pre-selected, which is based on historic information [9]. Therefore, such relays are not suitable for mobile scenarios. In addition, selecting malicious node as relay can make it is easy to launch any attack.

## 4. Discussion

Making the e-health systems efficient and affective can benefit the human society. Compromising a node could even result in loss of lives (humans or animals). As a summary, security is needed to be addressed at every level.

The security mechanism must be light weight, since we have limited resources. There are various kinds of attacks that can be launched in e-health systems. This could be passive i.e. regarding patients' data confidentiality etc., or active i.e., DoS etc. Issues related to DoS attack are explained below:

*4.1. Denial of Services*

Whenever the requested resource is not granted within due time, we can say DoS attack is occurred. DoS attack can active or passive. In former case the attacker launches attack, while in later case there is depletion of resources. Therefore, DoS attack is very critical to WBANs systems. We can avoid active attacks but better intelligence is required for mitigating the passive one such as lazy node behavior etc. We need to improve the performance of overall network in order to avoid such passive DoS attacks. That is why it can be very easy for adversary to launch DoS attack, as well as it can be launch by itself from inside of network. Therefore, it is very important to design better solutions for handling DoS, which is an open research area within e-health systems.

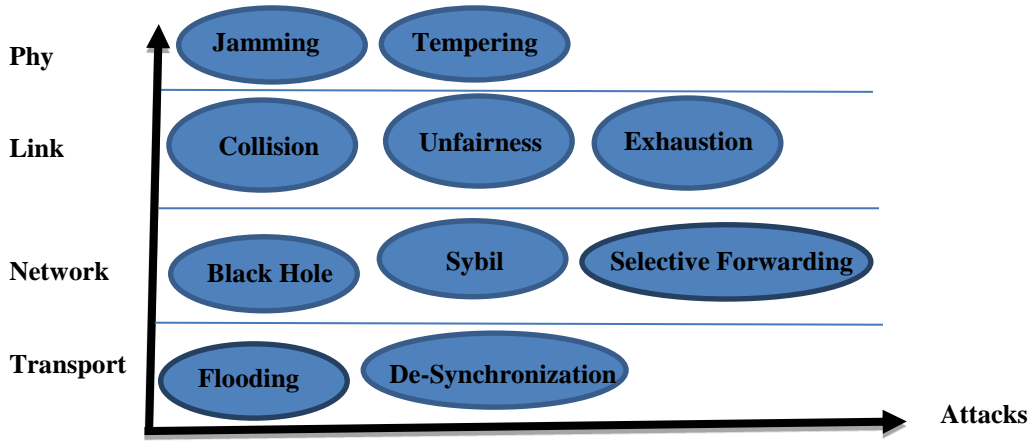

Fig. 3  DoS attacks on WBAN

All attacks we discussed in this paper are kind of active DoS attacks, as shown in Fig. 3. In WBANs the MAC layer is important, which deal with channel access and perform resource management. The depletion of resources and loss of connectivity leads to passive DoS attacks. One of the solutions in this regard is cooperation, as explained below.

*4.2. Cooperation*

As explained earlier, cooperation offers good solution for many problems in wireless networks. One of the main advantages is always-connected situation. With the cooperation we can ensure that patient is always connected even in case of disaster or loss of infrastructure.

Multiple relays assisting source-destination pair could allow higher diversity gain, which would increase the reliability. However, sending data frames via intermediate nodes are not always secure. The relay node may be malicious or its unavailability may lead to DoS itself. Hence, cooperation might itself become security risk.

In our opinion, there is need of secure cooperative networking protocol at MAC layer to enable cooperation in case of link failure. Such protocol must be reactive and based on local information. Such cooperation can be used to assist layer three-path re-computation (c.f. Fig 3.b).

*4.3. Our Findings*

- WBANs are prone to many attacks which results in packet loss, false redirection, modification which ultimately compromise confidentiality.
- Weak physical link in wireless communication subject to have high probability of to be compromised.
- An internal node can also be have malicious intent, so a trust has to been periodically been rechecked.
- Minimum level of centralized control in BWAN network has to be exercised to overcome single point failure.
- Scalability in WBAN has to carefully cater and optimal solution required generating alarm in case of problematic situation of denial of service attack.
- Eavesdropping in WBANs has to be critically addressed in communication which is apparently a major threat of reliable communication.
- Secure Route discovery and communication prompt researchers to an evolving issue for WBANS.

- The race between secure communication and attacks evolve as continuous rivalries. New security mechanism also restrains to have new type of attacks.
- DoS attack could be launched due to sacristy of resources.
- Resource management needs special attention in case of WBANs.
- Multiple copies of data via relays can increase reliability [13].
- Cooperation helps to improve overall throughput of systems and allow us to communicate via reliable nodes only.
- Distance and mobility can affect the data rate and channel quality, which could be mitigated via cooperation [12].

## 5. Conclusions and Future Work

This article provided a systematic overview of WBANs. WBAN provides monitoring of human health systems whether inside or outside. However, this is new technology and research is needed to address the conflicting challenges. In this paper, we discussed various issues related to WBANs. These issues are categorized as two main classes, i.e., Cooperation and Security issues.

Security issues need to be tackle at every layer and scenarios of E-health systems [10] [11]. Similarly, all the networks related issues such as mobility; QoS, energy, routing, distance etc., could be resolved via cooperation. Cooperation on the other hand adds additional threats, blockage and overhead.

Therefore, as a future work we aim to propose secure cooperative relaying solution for WBANs to address most of the inherit issues. Similarly, as a part of effective security mechanism we will work on various kind of DoS attack's detection and mitigations. In parallel we will devise a novel relay selection approach while relying over relays [14], [15].

## 6. Compliance with Ethical Standards:

*6.1. Conflict of Interest:*

- **Dr. Tauseef Jamal is Associate Professor at PIEAS University Islamabad, Pakistan.**

*6.2. Ethical approval:*

- **This article does not contain any studies with human participants or animals performed by any of the authors.**

.